\documentstyle[aps,twocolumn,tighten,epsfig]{revtex}

\begin{document}

\draft
\title{ HOW TO KICK A SOLITON ? }

\author{Jacek Dziarmaga\thanks{E-mail: {\tt J.P.Dziarmaga@durham.ac.uk}}
    and Wojtek Zakrzewski\thanks{E-mail: {\tt W.J.Zakrzewski@durham.ac.uk}} }

\address{Department of Mathematical Sciences,
         University of Durham, South Road, Durham, DH1 3LE,
         United Kingdom}
\date{June 25, 1997}
\maketitle
\tighten

\begin{abstract}
A simple method how to study response of solitons in dissipative systems
to external impulsive perturbations is developed. Thanks to nontrivial
choice of small parameter, the perturbative scheme captures genuine
nonlinear phenomena. The method is developed and tested by numerical
simulations for kinks in 1+1 dimensions and for skyrmions in 2+1
dimensions. Extension to models including second order time derivatives is
discussed.
\end{abstract}

  Let us consider a dissipative version of the $\phi^{4}$ theory in one
spatial dimension, which is defined, in appropriate dimensionless units,
by the nonlinear field equation

\begin{equation}\label{model}
\partial_{t}\phi=\partial_{x}^{2}\phi+2[1-\phi^{2}]\phi+\eta(t,x) \;\;.
\end{equation}
In the absence of noise, $\eta(t,x)=0$, Eq.(\ref{model}) admits static
kink solutions

\begin{equation}\label{kink}
\phi(t,x)=F(x)\equiv\tanh(x) \;\;.
\end{equation}
Antikinks are given by $F(-x)$. 

\subsection*{Spectrum of kink excitations.} Small perturbations around the
kink take the form $e^{-\gamma t}g(x)$.  Linearization of Eq.(\ref{model}) 
with respect to $g(x)$, for $\eta(t,x)=0$, gives

\begin{equation}\label{geq}
\gamma g(x)=-\frac{d^{2}}{dx^{2}}g(x)+
            [4-\frac{6}{\cosh^{2}(x)}]g(x) \;\;,
\end{equation}
which looks like a stationary Schr\"odinger equation for a potential well
problem. The eigenvalues and eigenstates can be tabulated as~\cite{jackiw}

\begin{eqnarray}\label{spectrum}
 \gamma,\;\;\;\;\;\;\;\;  && g(x)  \;\;,\nonumber\\
 0,\;\;\;\;\;\;\;\; && F'(x)=\frac{1}{\cosh^{2}(x)}  \;\;,\nonumber\\
 3,\;\;\;\;\;\;\;\; &&
                B(x)\equiv\frac{\sinh(x)}{\cosh^{2}(x)}\;\;,\nonumber\\
 4+k^{2}, && g_{k}(x)\equiv
 e^{ikx}[1+\frac{3ik\tanh(x)-3\tanh^{2}(x)}{1+k^2}]\;\;,
\end{eqnarray}
where $k$ is a real momentum. The zero mode ($\gamma=0$) is separated by a
gap from the first excited state ($\gamma=3$).

\subsection*{Naive perturbation theory (NPT).}

Let us consider the response of the kink to a solitary perturbation
$\eta(t,x)=\delta(t)\psi(x)$. Let the unperturbed state be
$\phi(t<0,x)=F(x)$; integration of Eq.(\ref{model}) over time in an
infinitezimal neighborhood of $t=0$ results in the initial configuration
$\phi(0^{+},x)=F(x)+\psi(x)$.  The function $\psi(x)$ can be expanded in
the basis (\ref{spectrum}) as $\psi(x)=\tilde{\psi}_{z}F'(x)+
\tilde{\psi}_{B}B(x)+ \int_{-\infty}^{\infty}dk\;\tilde{\psi}(k)g_{k}(x)$.
If $\psi(x)$ is sufficiently small, such that Eq.(\ref{model}) can be
linearized around $\phi(t,x)=F(x)$ for $t>0$, then the perturbed field is
well approximated by

\begin{eqnarray}
\phi(t>0,x)&\approx& 
F(x)+\tilde{\psi}_{z}F'(x)+e^{-3t}\tilde{\psi}_{B}B(x)+  \nonumber\\
&&\int_{-\infty}^{\infty}dk\; e^{-(4+k^{2})t}\tilde{\psi}(k)g_{k}(x) \;\;. 
\end{eqnarray}
After a sufficiently long time, as compared to the relaxation time of the
first excited state equal to $1/3$, the field relaxes to

\begin{eqnarray}
&&\phi(\infty,x)\approx F(x)+\tilde{\psi}_{z} F'(x)\approx 
F[x+\tilde{\psi}_{z}]\;,\nonumber\\
&&\tilde{\psi}_{z}=\frac{\int dx\;F'(x)\psi(x)}{\int dy\;F'(y)F'(y)} \;\;.
\end{eqnarray}
The kink is shifted to a new position, the shift is given by a projection
of the initial perturbation $\psi(x)$ onto the kink's translational zero
mode.  However plausible it may sound, this result is not very accurate
even for small perturbations. The reason is that the spectrum
(\ref{spectrum}) is relevant for perturbations around the initial kink at
the origin but it becomes less relevant as the kink drifts to its final
location at $-\tilde{\psi}_{z}$. The excited states in (\ref{spectrum}) 
are not orthogonal to the zero mode $F'[x+\tilde{\psi}_{z}]$ of the final
kink.

\subsection*{Improved perturbation theory (IPT).}

 An appropriate choice of perturbation theory may improve the accuracy.  Let
us consider again the single kick $\eta(t,x)=\delta(t)\psi(x)$ and let us
rewrite the initial perturbed field in the following form

\begin{equation}\label{balance}
\phi(0^{+},x)\equiv F(x)+\psi(x)=F(x-\xi^{+})+\mu(0^{+},x) \;\;.
\end{equation}
The above equation is just a tautology, unless we impose a constraint on
the function $\mu(0^{+},x)$. We require the function $\mu(0^{+},x)$ to be
orthogonal to the zero mode of the kink at $\xi^{+}$,

\begin{equation}\label{orto}
0=\int_{-\infty}^{+\infty}dx\; F'(x-\xi^{+})\mu(0^{+},x) \;\;.
\end{equation}
This orthogonality implies that $\mu(0^{+},x)$ can be expressed as a
combination of the excited states in the complete set (\ref{spectrum})
shifted to $\xi^{+}$,

\begin{equation}\label{expansion}
\mu(0^{+},x)=\tilde{\mu}_{B}\; B(x-\xi^{+})+
             \int_{-\infty}^{+\infty}dk\;\tilde{\mu}(k)\; 
                                         g_{k}(x-\xi^{+}) \;\;. 
\end{equation} 
So far we have made no approximation. The perturbative approximation is
defined by the assumption that $\mu(0^{+},x)$ is sufficiently small, so
that Eq.(\ref{model}) can be linearized in $\mu(t,x)$ around
$F(x-\xi^{+})$.  The linearized solution is given by

\begin{eqnarray}\label{memory}
\mu(t>0,x)&=&\tilde{\mu}_{B}e^{-3t}B(x-\xi^{+})+  \nonumber\\
&&\int_{-\infty}^{+\infty}dk\;\tilde{\mu}(k)\;e^{-(4+k^{2})t}g_{k}(x-\xi^{+}) 
\;\;. 
\end{eqnarray}
After a sufficiently long time $\mu(\infty,x)=0$ and the initial condition
on the R.H.S. of Eq.(\ref{balance}) evolves into $F(x-\xi^{+})$. Thus
$\xi^{+}$ is the final position of the kink after perturbation.

 To find $\xi^{+}$, let us project Eq.(\ref{balance}) on the zero mode
$F'(x-\xi^{+})$. Using Eq.(\ref{orto}) and the fact that $F(x)$ is
orthogonal to $F'(x)$, we obtain

\begin{equation}\label{tu}
\int^{+\infty}_{-\infty}dx\;F'(x-\xi^{+})[F(x)+\psi(x)]=0 
\;\;,
\end{equation}  
which can be further rewritten as

\begin{equation}\label{cond}
U(\xi^{+})+
\int^{+\infty}_{-\infty}dx\;\frac{\psi(x+\xi^{+})}{\cosh^{2}(x)}=0 \;\;,
\end{equation}
where $U$ is the smooth, antisymmetric and monotonic function

\begin{equation}\label{u}
U(\xi^{+})=
2[\frac{1}{\tanh(\xi)}-\frac{\xi}{\sinh^{2}(\xi)}] \;\;.
\end{equation}
Already from the general form of (\ref{cond}) it is clear, that $\xi^{+}$
is not linearly dependent on the magnitude of the perturbation $\psi(x)$. 
This response is nonlinear although we use a sort of perturbation theory.

\subsection*{Numerical simulation.}

Before we proceed with general discussion, let us consider some
quantitative examples and compare IPT predictions with direct numerical
simulations of Eq.(\ref{model}). We take the perturbation in the form

\begin{equation}\label{psi}
\psi(x)=-\frac{A}{\cosh^{2}(x-y)} \;\;,
\end{equation}
where $A$ is the amplitude and $y$ the localization of the perturbation.
Eq.(\ref{cond}) takes the form

\begin{equation}\label{A-ksi}
A=\frac{\sinh^{3}(\xi^{+}-y)
        [\frac{1}{\tanh(\xi^{+})}-\frac{\xi^{+}}{\sinh^{2}(\xi^{+})}]}
       {2[(\xi^{+}-y)\cosh(\xi^{+}-y)-\sinh(\xi^{+}-y)]} \;\;.
\end{equation}

  Fig.1 illustrates the dependence of the response $\xi^{+}$ on the
amplitude $A$ for a central perturbation with $y=0$. IPT works well up to
$A\approx 1$, or up to soliton shifts comparable with soliton size. Fig.2
shows the data for $A=0.2$ and $y$ varying in the range $-2<y<2$. If the
perturbation hits to the right of the kink, $y>0$, and is such that the
kink is driven to the right, $A>0$, the response is amplified because the
perturbed kink is flowing towards the source of the perturbation. This
positive feedback, which is missing in NPT, explains the asymetry in
Fig.2. 

  For some $y$ Eq.(\ref{A-ksi}) does not have a unique solution
$\xi^{+}=\xi^{+}(A)$ in the whole range of A. This usually happens for
$|A|>1$, when the perturbation theory is not reliable a priori. As an
example, Fig.3 illustrates the response for $y=2$ and $A>0$. The
theoretical function is not unique around $A\approx 1.3$. Numerical
results coarse grain over this singularity.  The perturbation (\ref{psi})
with $y=2$ creates a virtual antikink-kink pair just to the right of the
original kink (\ref{kink}). The result is a kink-antikink-kink
configuration.  For a weak perturbation ($A<1.3$) the antikink annihilates
with the right kink, leaving the original left kink shifted a bit to the
right. For $A>1.3$ the antikink annihilates the original left kink, the
net result is a huge shift to the right ($\xi^{+}>A$), much larger than
that expected from NPT. This is again the positive feedback of Fig.2 but
in a caricaturely enhanced form. This is a genuine nonlinear response and
it is amusing to find that IPT does capture its essential features.

\subsection*{Response to subsequent perturbations.}

$\xi^{+}$ is the position the kink relaxes to after a sufficiently long
time. Before the final equilibrium is reached, the function $\mu(t,x)$,
see Eq.  (\ref{memory}), carries information about the initial
perturbation. This information is lost after roughly $1/3$ units of time,
which is the relaxation time of the breather $B(x)$ (\ref{spectrum}). As
an example, let us consider the kink perturbed by the noise
   
\begin{equation}\label{noise}
\eta(t,x)=\delta(t)\psi(x)+\delta(t-T)\psi(x) \;\;,
\end{equation}
where $\psi(x)$ is given by (\ref{psi}) with, say, $y=0$ and $A>0$. At the time 
$t=T$ the field is hit for the second time,

\begin{eqnarray}
\phi(T^{+},x)&=&F(x-\xi^{+})+\mu(T,x)+\psi(x)=        \nonumber\\
             && F(x-\xi^{++})+\bar{\mu}(T^{+},x) \;\;,
\end{eqnarray}
while it still remembers the first perturbation $\mu(T,x)$.  Repeating
steps leading to (\ref{cond}), one obtains

\begin{equation}\label{plusplus}
U(\xi^{++}-\xi^{+})+\int_{-\infty}^{+\infty}dx\;
\frac{\psi(x+\xi^{++})+\mu(T,x+\xi^{++})}
     {\cosh^{2}(x)}=0 \;\;.
\end{equation}
There are two limiting cases. If $T=0$ the two kicks add up to a solitary
perturbation $2\psi(x)$. If the function $\xi^{+}(A,y)$ is a solution of
Eq.(\ref{A-ksi}), then the final shift of the kink can be expressed as
$\xi^{++}(T=0)=\xi^{+}(2A,0)$. In the limit $T=\infty$, the kink has
enough time to settle down at $\xi^{+}(A,0)$ after the first perturbation
before it is shifted by the second kick to
$\xi^{++}(T=\infty)=\xi^{+}(A,0)+\xi^{+}[A,-\xi^{+}(A,0)]$.  The two
shifts are different, $\xi^{++}(T=\infty)>\xi(T=0)$. Fig.4 gives numerical
results for various values of $T$. The limit $\xi^{++}(T=\infty)$ is
essentially achieved for $T$ greater than the relaxation time of the
breather mode (\ref{spectrum}), $T>1/3$. For shorter intervals $T$ the
memory function $\mu(T,x)$ in Eq.(\ref{plusplus}) cannot be neglected. 

  The second kick probes the position of the kink at the time $T$ after
the first perturbation. Roughly speaking, the kink remains unshifted for
$T<1/3$, it jumps to $\xi^{+}(A,0)$ for $T>1/3$. The $\delta(t)$-like time
shape of our impulse may be good approximation for impulses lasting less
than $1/3$.

\subsection*{ Higher order time derivatives. }

 IPT can be generalized to models with higher order time derivatives. For
example, a modification
 
\begin{equation}
\partial_{t}\phi\rightarrow
\partial_{t}\phi+\alpha\partial^{2}_{t}\phi
\end{equation}
on the LHS of Eq.(\ref{model}) changes the initial conditions after the
perturbation $\eta(t,x)=\delta(t)\psi(x)$ to:
$\phi(0^{+},x)=F(x)\;,\;\partial_{t}\phi(0^{+},x)=\psi(x)$. The perturbed
field can be written as
$\phi(t>0,x)=F(x-\xi^{+})+(A_{1}+A_{2}e^{-t/\alpha})F'(x-\xi^{+})+R(t,x)$.
$R(t,x)$ can be expanded in the excited states which decay with time. For
$\xi^{+}$ to be the final kink position, $\phi(\infty,x)=F(x-\xi^{+})$, we
must demand $A_{1}=0$. This constraint plus the initial conditions result
in the following equation for $\xi^{+}$
  
\begin{equation}\label{alpha}
U(\xi^{+})+\alpha
\int^{+\infty}_{-\infty}dx\;\frac{\psi(x+\xi^{+})}{\cosh^{2}(x)}=0 \;\;,
\end{equation}
which is very similar to Eq.(\ref{cond}). It is important to realize that
the $\alpha\rightarrow 0$ limit of Eq.(\ref{alpha}) does not give
Eq.(\ref{cond}), which holds for $\alpha=0$. The more general model is
discontinuous at $\alpha=0$. 
 
  We have performed several numerical simulations to test
Eq.(\ref{alpha}).  The accuracy is the same as for $\alpha=0$.

\subsection*{IPT and skyrmions in 2+1 dimensions.}

  Once IPT has passed the tests for kinks, it would be interesting to
check how it works in higher dimensional models. Let us consider the
diffusion equation in the $\vec{x}=(x_{1},x_{2})$ plane

\begin{equation}\label{diff}
\partial_{t}\hat{M}=
P_{\hat{M}}[-\frac{\delta W}{\delta\hat{M}}
                     +\hat{\eta}(t,\vec{x})]\;\;,
\end{equation}
where $\hat{M}(t,\vec{x})$ is a unit ($\hat{M}\hat{M}=1$) magnetization
vector, $P_{\hat{M}}$ is a projector on the subspace orthogonal to
$\hat{M}$, $W=\int d^{2}x\;w$ is the static energy functional with the
energy density

\begin{eqnarray}\label{w}
w&=&\frac{1}{2}\partial_{k}\hat{M}\partial_{k}\hat{M}+
    \frac{1}{2}a_{1}( M_{1}^2+M_{2}^2 )+    \nonumber\\
&&\frac{1}{4}a_{2}[ (\partial_{k}\hat{M}\partial_{k}\hat{M})^{2}-
                    (\partial_{k}\hat{M}\partial_{l}\hat{M})^{2} ]
\end{eqnarray}
being a sum of the exchange, easy axis anisotropy and Skyrme energy
densities respectively. The energy is minimized by the skyrmion
configuration~\cite{wjz-np}

\begin{equation}\label{skyrmion}
\hat{F}(x_{1},x_{2})=
\{ \sin[q(r)]\cos\theta, \sin[q(r)]\sin\theta, \cos[q(r)] \}
\end{equation}
where $(r,\theta)$ are polar coordinates in the $(x_{1},x_{2})$ plane and
$q(r)$ is a profile such that $q(0)=\pi$ and $q(\infty)=0$. $q(r)$
has been determined by a shooting method for $a_{1}=1$ and $a_{2}=0.1$,
the result is shown in Fig.5. 

  The initial field after the perturbation
$\hat{\eta}(t,\vec{x})=\delta(t)\hat{\psi}(\vec{x})$ is a bit complicated
because of the projector on the RHS of Eq.(\ref{diff}).  At any point
$\vec{x}$ the integration of Eq.(\ref{diff}) around $t=0$ gives the
initial perturbed magnetization

\begin{eqnarray}\label{mplus}
&& c_{-}\equiv \frac{\hat{\psi}}{|\hat{\psi}|}\hat{F} \;\;,\;\; 
   c_{+} \equiv \frac{c_{-}+\tanh(|\hat{\psi}|)}
                     {1+c_{-}\tanh(|\hat{\psi}|)} \;\;,      \nonumber\\
&& \hat{M}(t=0^{+})=
   c_{+}\frac{\hat{\psi}}{|\hat{\psi}|}+
   \sqrt{\frac{1-c_{+}^{2}}{1-c_{-}^{2}}}\;
   P_{\frac{\hat{\psi}}{|\hat{\psi}|}}[\hat{F}] \;\;.       
\end{eqnarray}
Similarly as for kinks, compare Eqs.(\ref{balance},\ref{tu}), the final
position $\vec{\xi}^{+}$ of the skyrmion can be found from the condition
that the initial perturbed field is orthogonal to the $k=1,2$
translational zero modes of the final skyrmion

\begin{equation}\label{condition}
0=\int d^{2}x\;
  \hat{M}(0^{+},\vec{x}) 
  \frac{\partial}{\partial x_{k} }
  \hat{F}(\vec{x}-\vec{\xi}^{+})\;\;.
\end{equation}

  We performed numerical simulations of Eq.(\ref{diff}) for the kick
$\hat{\psi}(\vec{x})=-0.3 \frac{\partial}{\partial x_{1}
}\hat{F}(\vec{x}-\vec{y})$ with $\vec{y}=(y_{1},0)$ restricted to the
$x_{1}$-axis. As a result of the perturbation the skyrmion was shifted to
a new position at $\vec{\xi}^{+}=(\xi^{+}_{1},0)$. The function
$\xi^{+}_{1}(y_{1})$ is presented in Fig.6. The maximum of the function is
shifted to the right and the minima are shifted to the left as a result of
similar positive feedback as in Fig.2. 

  IPT works well for skyrmions in $2+1$ dimensions. The model (\ref{diff})
can be generalized to, say, Landau-Lifshitz equation with dissipation.
Modification of the LHS of Eq.(\ref{diff}), 
$\partial_{t}\hat{M}\rightarrow\partial_{t}\hat{M}+
 \lambda \hat{M}\times\partial_{t}\hat{M}$, changes
$\hat{M}(0^{+},\vec{x})$ but the condition (\ref{condition}) remains the
same.

\subsection*{Remarks.} 

The theory outlined above can be upgraded to a fully fledged theory of
soliton diffusion driven by a random noise. Another possible line of
development is a generalization to the case of dissipative
multisolitons~\cite{jd}.

\acknowledgements

J.D. would like to thank Nick Manton for a comment which initiated his
work on this project. J.D. was supported by a UK PPARC grant.

\begin{figure}
\begin{center}
\epsfig{file=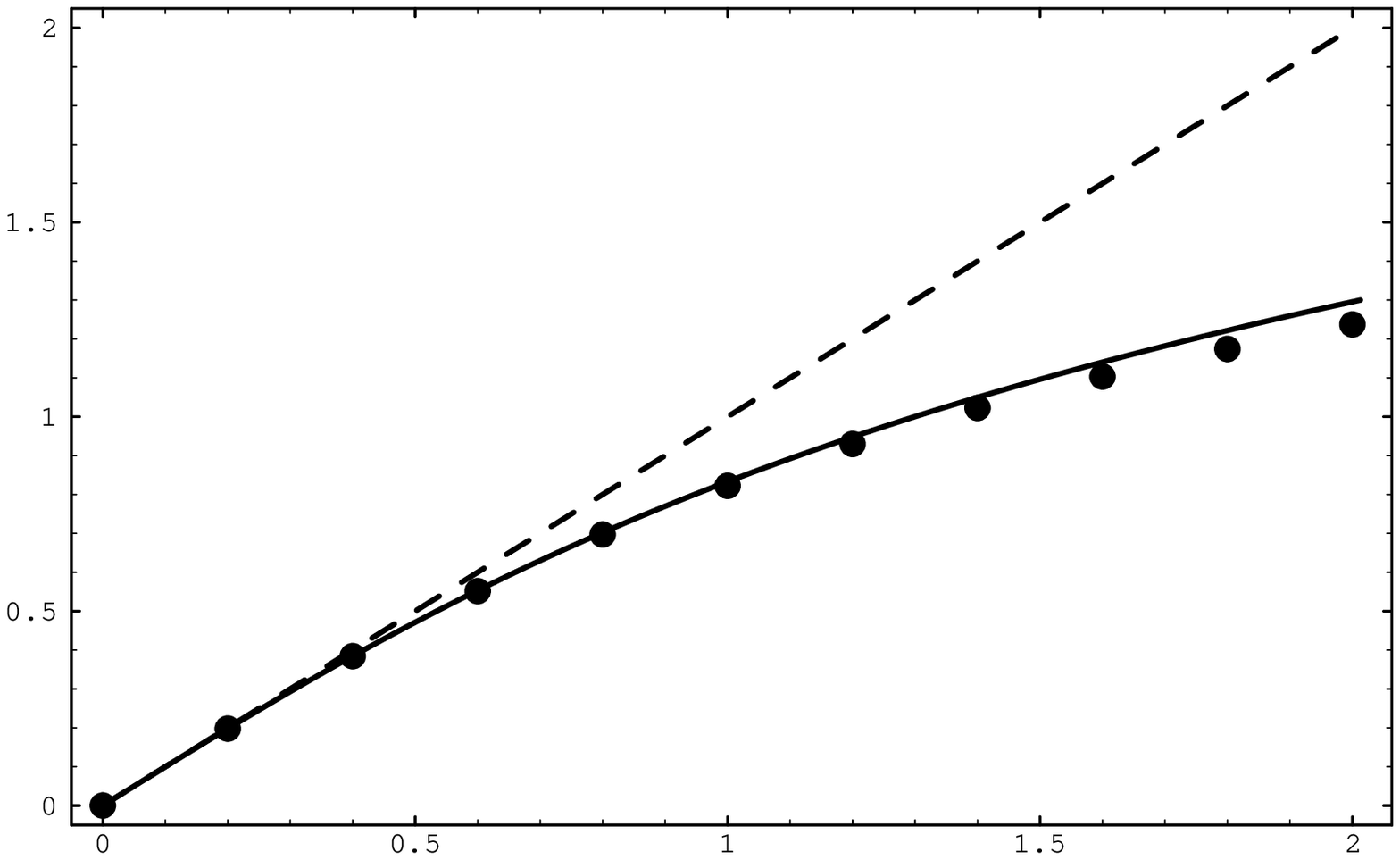,width=8cm}
\end{center}
\end{figure}

FIGURE 1. $\xi^{+}$ for the kink as a function of $A$ for $y=0$.  Dashed
line - NPT, solid line - IPT and dots - numerical simulation. 

\begin{figure}
\begin{center}
\epsfig{file=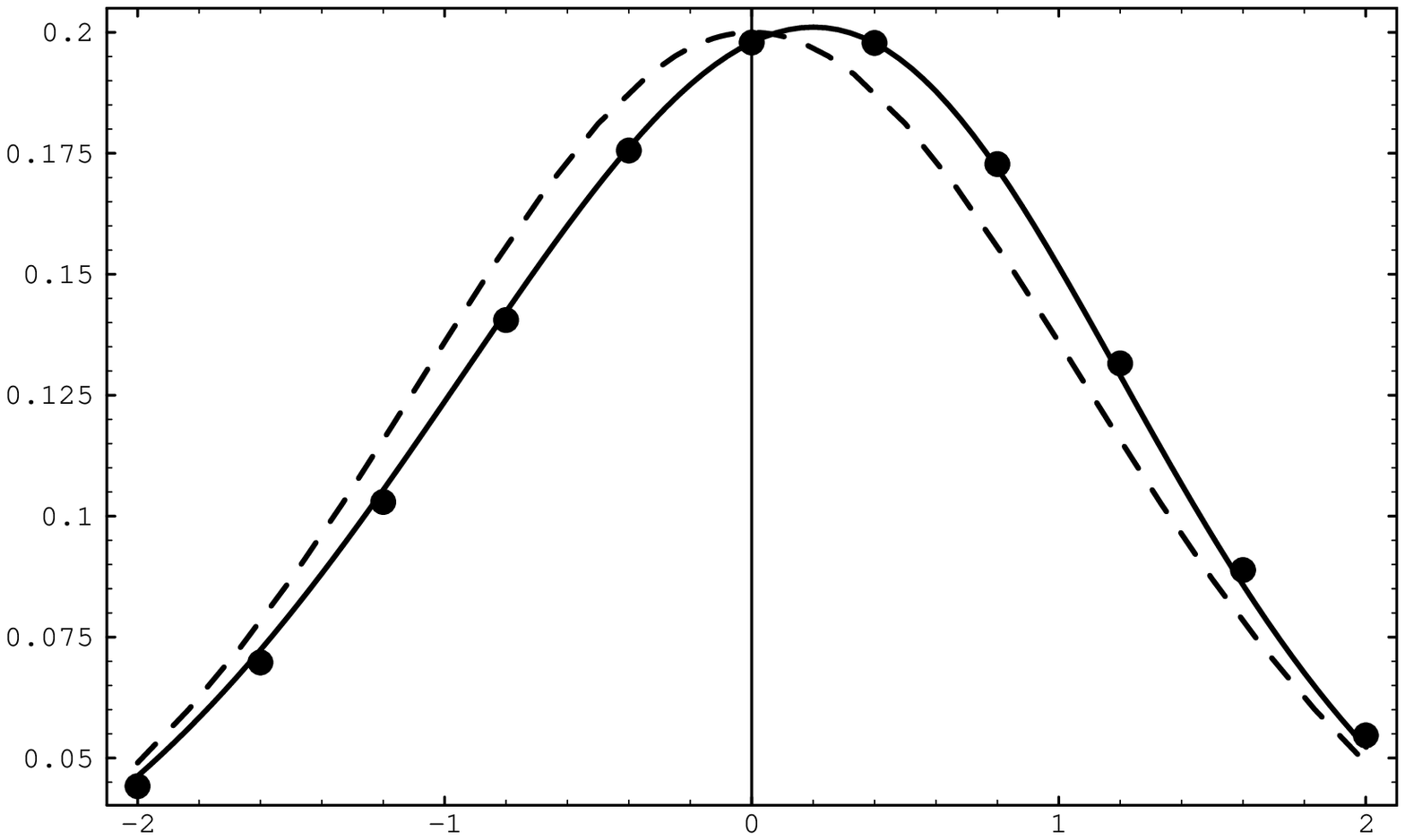,width=8cm}
\end{center}
\end{figure}
 
FIGURE 2. $\xi^{+}$ for the kink as a function of $y$ for $A=0.2$.
 
\begin{figure}
\begin{center}
\epsfig{file=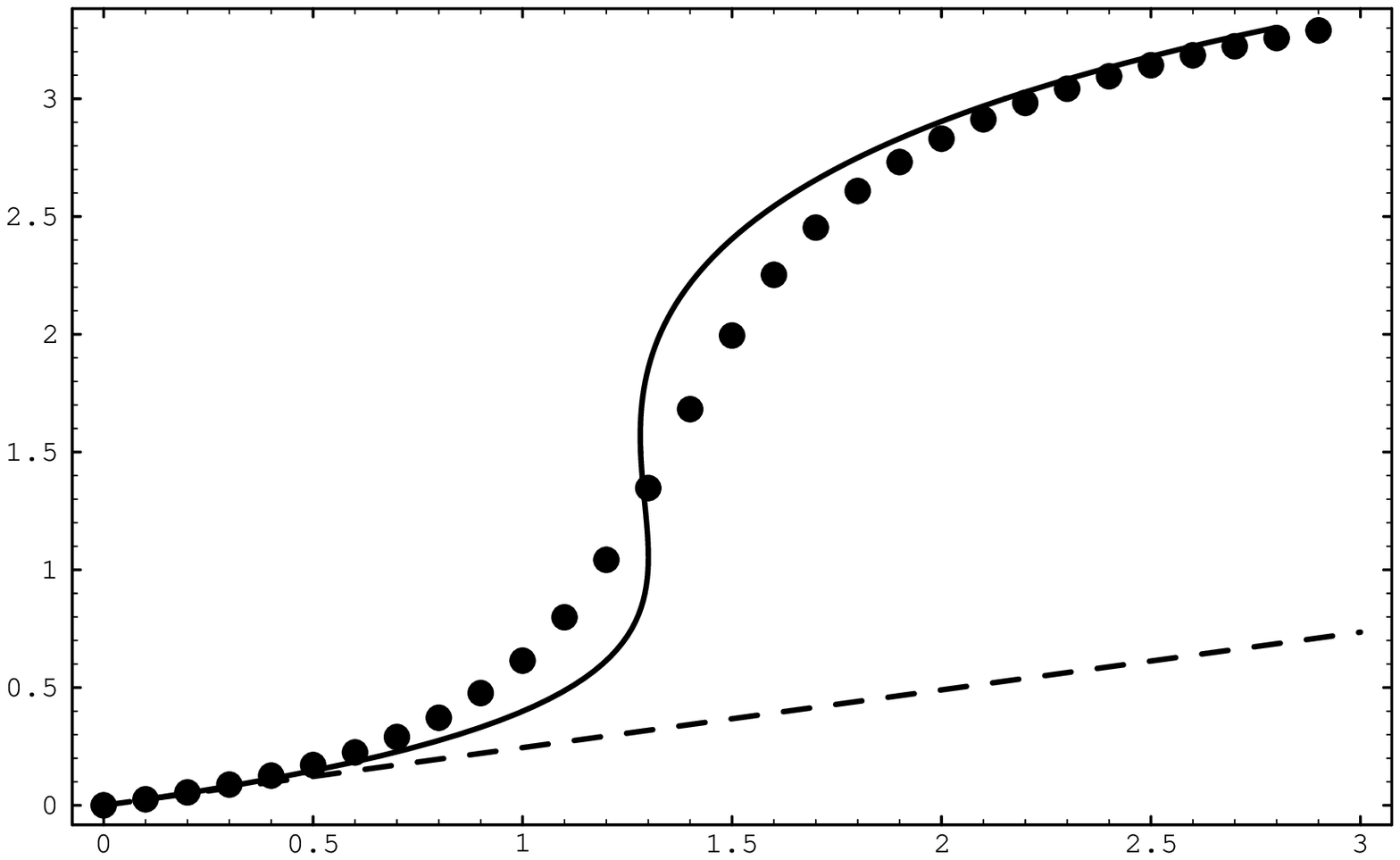,width=8cm}
\end{center}
\end{figure}

FIGURE 3. $\xi^{+}$ for the kink as a function of $A$ for $y=2$.
 
\begin{figure}
\begin{center}
\epsfig{file=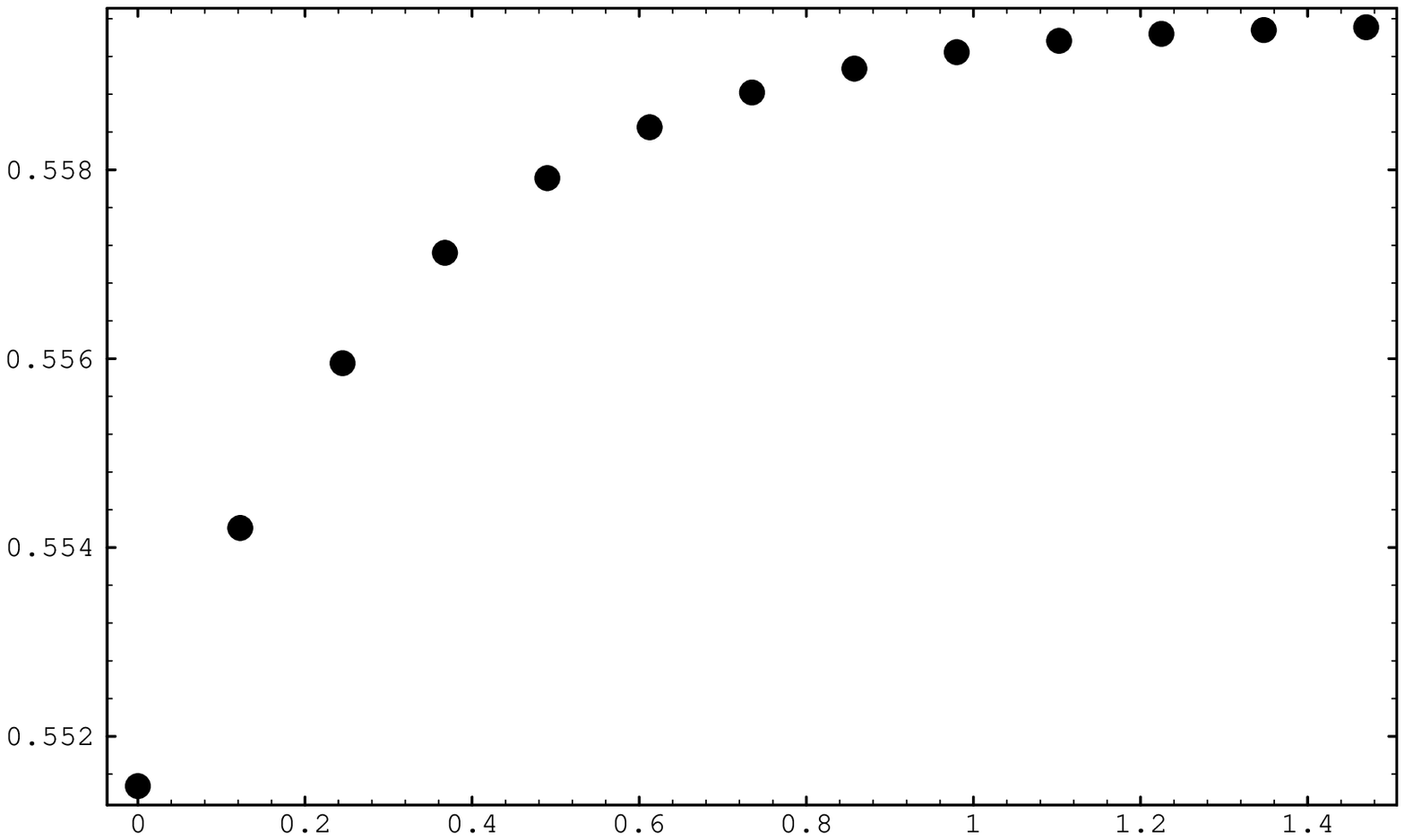,width=8cm}
\end{center}
\end{figure}

FIGURE 4. Numerical simulation results for $\xi^{++}$ after two subsequent
perturbations separated by $T$ as a function of $T$. Amplitude A=0.3. 

\begin{figure}
\begin{center}
\epsfig{file=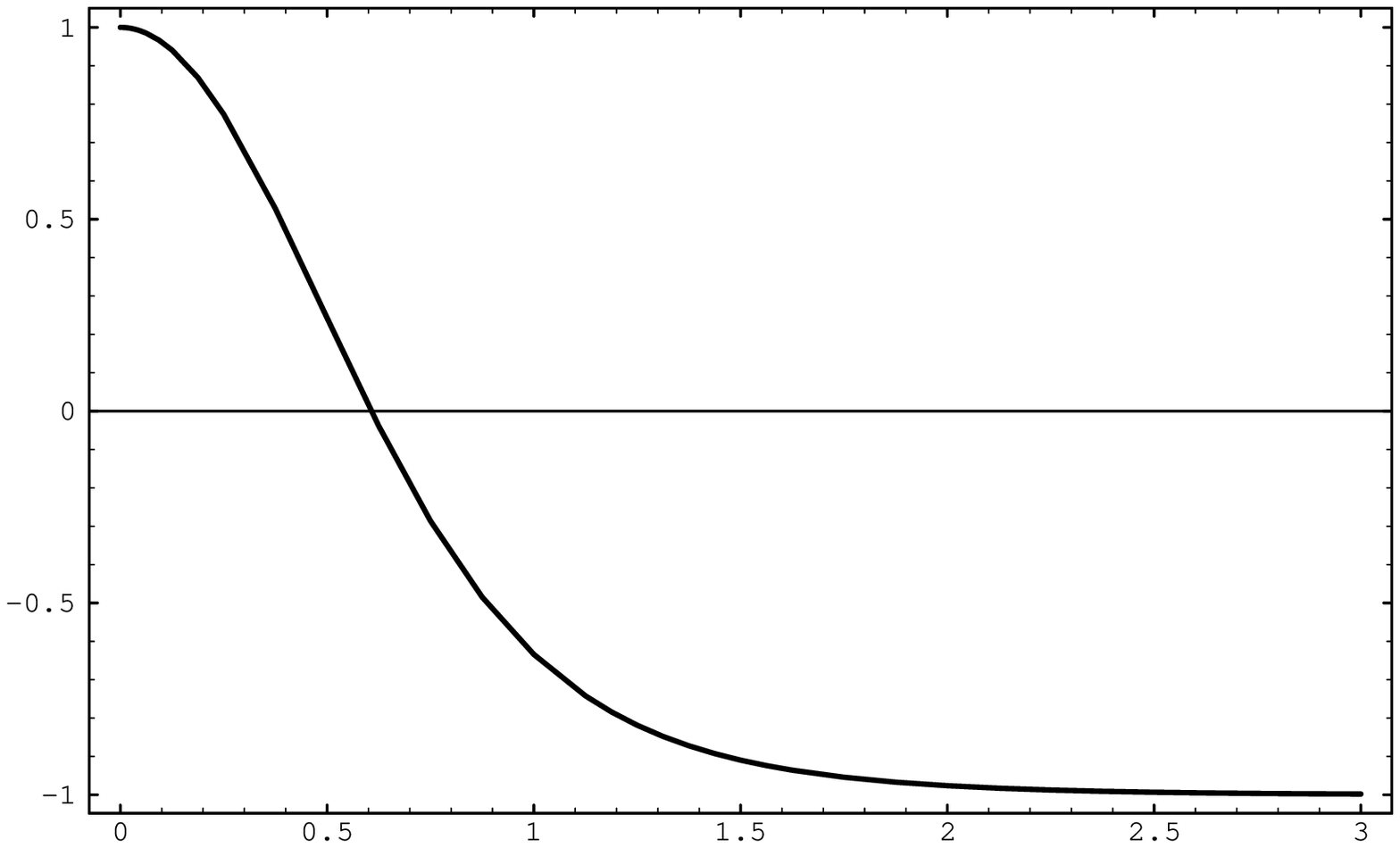,width=8cm}
\end{center}
\end{figure}
 
FIGURE 5. The magnetization $M_{3}=\cos[q(r)]$ as a function of $r$.

\begin{figure}
\begin{center}
\epsfig{file=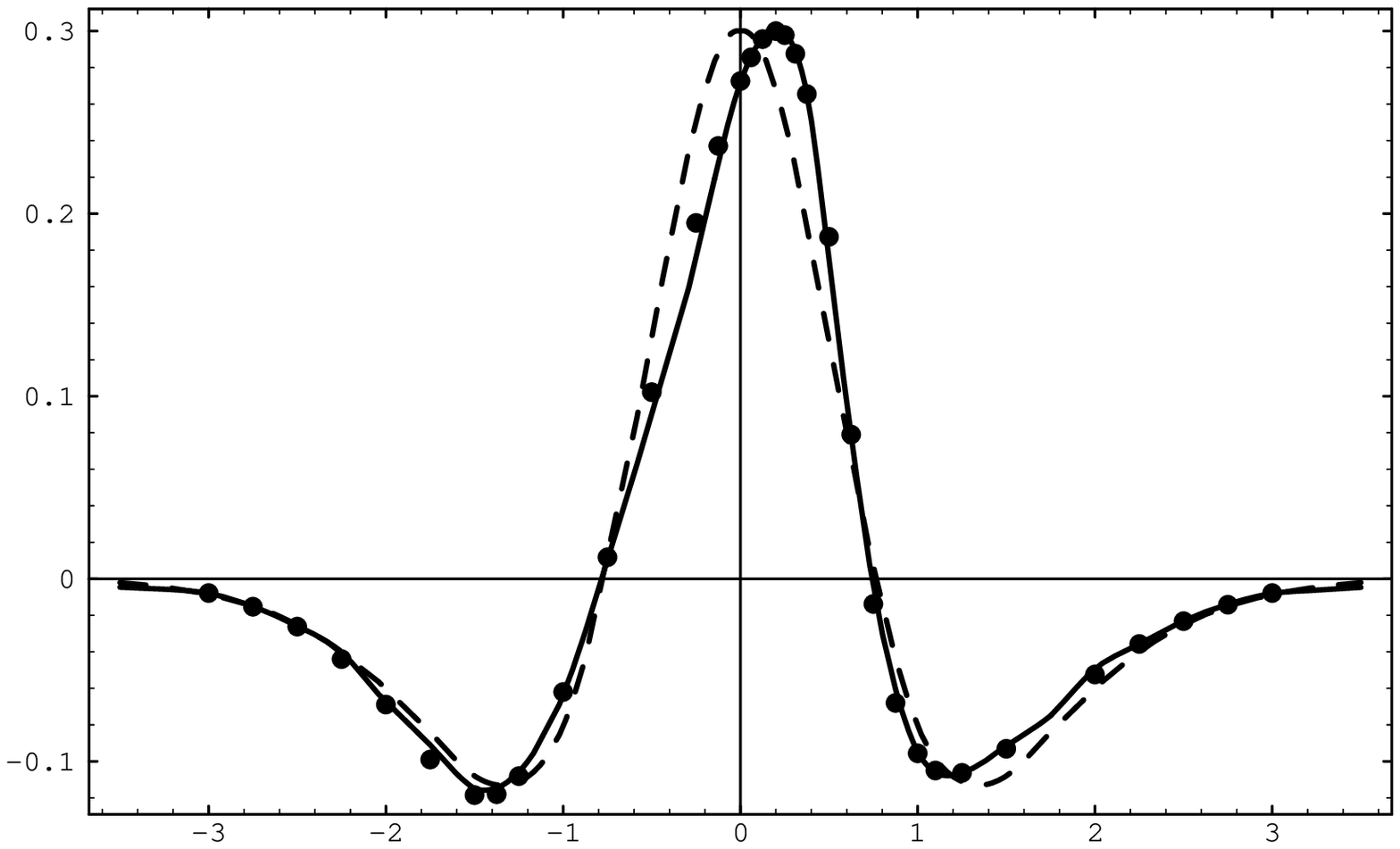,width=8cm}
\end{center}
\end{figure}
 
FIGURE 6. $\xi^{+}_{1}$ for the skyrmion as a function of $y_{1}$ for
$A=0.3$. Dashed line - NPT, solid line - IPT and dots - numerical
simulation. 

\end{document}